\documentstyle[12pt]{article}

\catcode`@=11
\newcount\@tempcntc
\def\@citex[#1]#2{\if@filesw\immediate\write\@auxout{\string\citation{#2}}\fi
  \@tempcnta\z@\@tempcntb\m@ne\def\@citea{}\@cite{\@for\@citeb:=#2\do
    {\@ifundefined
       {b@\@citeb}{\@citeo\@tempcntb\m@ne\@citea\def\@citea{,}{\bf
?}\@warning
       {Citation `\@citeb' on page \thepage \space undefined}}%
    {\setbox\z@\hbox{\global\@tempcntc0\csname b@\@citeb\endcsname\relax}%
     \ifnum\@tempcntc=\z@ \@citeo\@tempcntb\m@ne
       \@citea\def\@citea{,}\hbox{\csname b@\@citeb\endcsname}%
     \else
      \advance\@tempcntb\@ne
      \ifnum\@tempcntb=\@tempcntc
      \else\advance\@tempcntb\m@ne\@citeo
      \@tempcnta\@tempcntc\@tempcntb\@tempcntc\fi\fi}}\@citeo}{#1}}
\def\@citeo{\ifnum\@tempcnta>\@tempcntb\else\@citea\def\@citea{,}%
  \ifnum\@tempcnta=\@tempcntb\the\@tempcnta\else
   {\advance\@tempcnta\@ne\ifnum\@tempcnta=\@tempcntb \else
\def\@citea{--}\fi
    \advance\@tempcnta\m@ne\the\@tempcnta\@citea\the\@tempcntb}\fi\fi}
\catcode`@=12
\begin{document}
\title{\vskip-3cm{\baselineskip14pt
\centerline{\normalsize DESY 00-047\hfill ISSN 0418-9833}
\centerline{\normalsize hep-ph/0005060\hfill}
\centerline{\normalsize May 2000\hfill}}
\vskip1.5cm
Gauge-Independent $W$-Boson Partial Decay Widths}
\author{{\sc Bernd A. Kniehl, Fantina Madricardo, Matthias Steinhauser}\\
{\normalsize II. Institut f\"ur Theoretische Physik, Universit\"at
Hamburg,}\\
{\normalsize Luruper Chaussee 149, 22761 Hamburg, Germany}}

\date{}

\maketitle

\thispagestyle{empty}

\begin{abstract}
We calculate the partial decay widths of the $W$ boson at one loop in the
standard model using the on-shell renormalization scheme endowed with a
gauge-independent definition of the Cabibbo-Kobayashi-Maskawa (CKM) mixing
matrix.
We work in $R_\xi$ gauge and explicitly verify that the final expressions are
independent of the gauge parameters.
Furthermore, we establish the relationship between the on-shell and
$\overline{\mathrm{MS}}$ definitions of the CKM matrix, both in its generic
form and in the Wolfenstein parameterization.
As a by-product of our analysis, we recover the beta function of the CKM
matrix.

\medskip

\end{abstract}

\newpage

\section{Introduction}

The discovery of the $W$ boson at the CERN S$p\bar p$S collider in 1983
\cite{Arn83} was one of the great successes of the standard model (SM) of the
electroweak interactions.
The properties of the $W$ boson, including its partial decay widths, have been
extensively studied at the CERN S$p\bar p$S collider and the Fermilab
Tevatron, and since 1995 also at the CERN Large Electron-Positron Collider
(LEP2).
On the theoretical side, the one-loop QED and electroweak radiative
corrections to the partial decay widths of the $W$ boson were calculated in
the light-fermion approximation in Refs.~\cite{Mar73,Ino80}, respectively, and
for finite fermion masses in Ref.~\cite{Den90}.
The QCD corrections, which are present for the hadronic decay modes, were 
computed at one loop for arbitrary quark masses in Refs.~\cite{Den90,Cha86}.
The two- and three-loop QCD corrections for massless quarks may be
extracted from Ref.~\cite{Gor88}, and the respective terms proportional to
$m_q^2/M_W^2$, where $m_q$ is a generic quark mass, may be found in
Ref.~\cite{Che97}.

In the calculation of the electroweak corrections, the treatment of the
Cabibbo-Kobaya\-shi-Maskawa (CKM) mixing matrix \cite{Cab63}, which rotates the
weak eigenstates of the quark fields into their mass eigenstates, deserves
special attention.
In the approximation of neglecting the down-quark masses against the $W$-boson
mass, the CKM matrix can be taken to be unity, so that it does not need to be
renormalized.
In fact, this avenue was taken in Refs.~\cite{Ino80,Den90}.
As a matter of principle, however, the CKM matrix elements must be
renormalized because they are parameters of the bare Lagrangian.
This was realized for the Cabibbo angle in the SM with two fermion generations
in a pioneering paper by Marciano and Sirlin \cite{Mar75}.
A compact and plausible on-shell renormalization prescription for the CKM
matrix of the three-generation SM was proposed in Ref.~\cite{Den90a} on the
basis of a detailed inspection of the ultraviolet singularities that are left
over in the one-loop expressions for the hadronic decay widths of the $W$
boson if one does not include renormalization constants for the CKM matrix
elements appearing in the tree-level formulas.
Since the analysis of Ref.~\cite{Den90a} was performed in a specific gauge,
namely in 't~Hooft-Feynman gauge, it could not be checked if the final
one-loop results are gauge independent as required.
A recent analysis in $R_\xi$ gauge has revealed that this is not the case
\cite{Mad99,GGM}.
In fact, the finite parts of renormalization constants for the CKM matrix
elements as defined in Ref.~\cite{Den90a} do depend on the gauge parameter
$\xi_W$ associated with the $W$ boson, and so do the renormalized CKM matrix
elements.
In Refs.~\cite{Mad99,GGM}, an alternative renormalization prescription for the
CKM matrix was proposed which, at one loop, is consistent with the relevant
Ward-Takahashi identities and avoids this problem.

In this paper, we explicitly calculate the $W$-boson partial decay widths at
one loop in the on-shell renormalization scheme supplemented by the
alternative renormalization prescription for the CKM matrix \cite{Mad99,GGM}.
For simplicity, we henceforth refer to this scheme as the on-shell scheme.
As in Ref.~\cite{Den90}, we keep the full dependence on the fermion masses.
We work in $R_\xi$ gauge, with arbitrary gauge parameters $\xi_W$, $\xi_Z$, 
and $\xi_A$, and verify that the final results are indeed independent of them.
Using up-to-date information on the input parameters, we present quantitative
predictions for the various leptonic and hadronic decay widths of the $W$ 
boson, which can be readily confronted with precise experimental data from the
Tevatron and from LEP2.
Furthermore, we establish the relationships between the CKM matrix elements 
renormalized according to the modified minimal subtraction
($\overline{\mathrm{MS}}$) scheme and their counterparts in the on-shell
scheme.
We also provide similar relationships appropriate for the Wolfenstein
parameterization.
From these relationships, we recover the beta functions of the CKM matrix
elements \cite{Sas86}, which may be relevant for studies within grand unified
theories.

This paper is organized as follows.
In Sec.~\ref{sec:two}, we establish our formalism and outline our calculation.
In Sec.~\ref{sec:three}, we exhibit the relationships between the CKM matrix
elements defined in the on-shell and $\overline{\mathrm{MS}}$ schemes and
extract their beta functions.
In Sec.~\ref{sec:four}, we present our quantitative predictions.
Our conclusions are contained in Sec.~\ref{sec:five}.
In the Appendix, we list the fermion two-point functions at one loop in
$R_\xi$ gauge.

\section{Analytical Results\label{sec:two}}

We now describe our analytical analysis, largely adopting the notations from
Ref.~\cite{DenHabil}.
We consider the two-particle decay of the $W^+$ boson to generic leptons or
quarks,
\begin{equation}
W^+(k)\to f_i(p_1)\bar f_j^\prime(p_2),
\label{eq:pro}
\end{equation}
where $f_i=\nu_e,\nu_\mu,\nu_\tau,u,c$, $f_j^\prime=e,\mu,\tau,d,s,b$, the bar
indicates antiparticles, and the four-momenta are specified in parentheses.
We wish to calculate the electroweak and QCD one-loop corrections to the 
partial decay width of process~(\ref{eq:pro}), with finite fermion masses
$m_{f,i}$ and $m_{f^\prime,j}$ and general CKM matrix $V_{ij}$.
The result for the charge-conjugate process, $W^-\to \bar f_if_j^\prime$, will
be the same.
The relevant standard matrix elements read \cite{DenHabil}
\begin{eqnarray}
{\cal M}_1^\sigma&=&\bar u(p_1)\!\!\not\!\varepsilon(k)\omega_\sigma v(p_2),
\nonumber\\
{\cal M}_2^\sigma&=&\bar u(p_1)\omega_\sigma v(p_2)\varepsilon(k)\!\cdot\!p_1,
\end{eqnarray}
where $\omega_\pm=(1\pm\gamma_5)/2$, $\varepsilon(k)$ is the polarization 
four-vector of the $W^+$ boson, and $\bar u(p_1)$ and $v(p_2)$ are the spinors
of the fermions $f_i$ and $\bar f_j^\prime$, respectively.
It is convenient to define~\cite{DenHabil}
\begin{eqnarray}
G_1^-&=&\sum_{\rm pol}{\cal M}_1^{-\dagger}{\cal M}_1^-
=2M_W^{2}-m_{f,i}^2-m_{f^\prime,j}^2
-\frac{\left(m_{f,i}^2-m_{f^\prime,j}^2\right)^2}{M_W^2},
\nonumber\\
G_1^+&=&\sum_{\rm pol}{\cal M}_1^{-\dagger}{\cal M}_1^+
=6m_{f,i}m_{f^\prime,j},
\nonumber\\ 
G_2^-&=&\sum_{\rm pol}{\cal M}_1^{-\dagger}{\cal M}_2^-
=-\frac{m_{f,i}}{2}\,
\frac{\kappa^2\left(M_W^2,m_{f,i}^2,m_{f^\prime,j}^2\right)}{M_W^2},
\nonumber\\
G_2^+&=&\sum_{\rm pol}{\cal M}_1^{-\dagger}{\cal M}_2^+
=-\frac{m_{f^\prime,j}}{2}\,
\frac{\kappa^2\left(M_W^2,m_{f,i}^2,m_{f^\prime,j}^2\right)}{M_W^2},
\label{eq:ms}
\end{eqnarray}
where $\kappa(x,y,z)=\sqrt{x^2+y^2+z^2-2(xy+yz+zx)}$ is the K\"all\'en
function and it is\break
summed over the polarization states of the $W^+$ boson and
the spin states of the fermions $f_i$ and $f^\prime_j$.
Let us first deal with the quark case, which is more involved.
In the Born approximation, the transition ($T$) matrix element of
process~(\ref{eq:pro}) then reads
\begin{equation}
{\cal M}_0^{Wf_if_j^\prime}=-\frac{eV_{ij}}{\sqrt2s_w}{\cal M}_1^-,
\label{eq:mlo}
\end{equation}
where $e$ is the electron charge magnitude and $s_w^2=1-c_w^2=1-M_W^2/M_Z^2$.
Thus, the partial decay width is given by
\begin{equation}
\Gamma_0^{Wf_if_j^\prime}=\frac{N_C^f\alpha|V_{ij}|^2}{24s_w^2M_W^3}
\kappa\left(M_W^2,m_{f,i}^2,m_{f^\prime,j}^2\right)G_1^-,
\label{eq:glo}
\end{equation}
where $N_C^f=3$ and $\alpha=e^2/(4\pi)$ is Sommerfeld's fine-structure
constant.

The one-loop-corrected $T$-matrix element of process~(\ref{eq:pro}) emerges
from Eq.~(\ref{eq:mlo}) by including the renormalization constants for the
parameters $e$, $s_w$, and $V_{ij}$, those for the $W^+$, $f_i$, and
$f_j^\prime$ fields, and the proper vertex correction.
It reads
\begin{eqnarray}
{\cal M}_1^{Wf_if_j^\prime}&=&-\frac{eV_{ij}}{\sqrt2s_w}\left\{
{\cal M}_1^-\left[1+\frac{\delta e}{e}-\frac{\delta s_w}{s_w}
+\frac{\delta V_{ij}}{V_{ij}}+\frac{1}{2}\delta Z_W
+\frac{1}{2}\sum_k\left(\delta Z_{ik}^{f,L\dagger}V_{kj}
+V_{ik}\delta Z_{kj}^{f^\prime,L}\right)\right]
\right.\nonumber\\
&&{}+\left.
\sum_{a=1}^{2}\sum_{\sigma=\pm}{\cal M}_a^\sigma
\delta F_a^\sigma\left(M_W,m_{f,i},m_{f^\prime,j}\right)\right\},
\label{eq:mol}
\end{eqnarray}
where $\delta F_a^\sigma$ are electroweak form factors.
The various renormalization constants may be expressed in terms of the
unrenormalized, one-particle-irreducible two-point functions of the gauge
bosons and fermions, as
\begin{eqnarray}
\frac{\delta e}{e}&=&-\frac{1}{2}\left(\frac{s_w}{c_w}\delta Z_{ZA}
+\delta Z_{AA}\right)
\nonumber\\
&=&-\frac{s_w}{c_w}\,\frac{\Pi^{ZA}(0)}{M_Z^2}
+\frac{1}{2}\left.\frac{\partial\Pi^{AA}(k^2)}{\partial k^2}\right|_{k^2=0},
\label{eq:de}\\
\frac{\delta s_w}{s_w}&=&-\frac{c_w^2}{2s_w^2}\left(
\frac{\delta M_W^2}{M_W^2}-\frac{\delta M_Z^2}{M_Z^2}\right)
\nonumber\\
&=&-\frac{c_w^2}{2s_w^2}\widetilde{\mbox{Re}}\left(
\frac{\Pi^{WW}\left(M_W^2\right)}{M_W^2}
-\frac{\Pi^{ZZ}\left(M_Z^2\right)}{M_Z^2}\right),
\label{eq:ds}\\
\delta V_{ij}&=&\frac{1}{2}\left\{
\sum_{k\ne i}\frac{m_{f,i}^2+m_{f,k}^2}{m_{f,i}^2-m_{f,k}^2}
\left[\Sigma_{ik}^{f,L}(0)+2\Sigma_{ik}^{f,S}(0)\right]V_{kj}
\right.\nonumber\\
&&{}-\left.
\sum_{k\ne j}V_{ik}
\frac{m_{f^\prime,k}^2+m_{f^\prime,j}^2}{m_{f^\prime,k}^2-m_{f^\prime,j}^2}
\left[\Sigma_{kj}^{f^\prime,L}(0)+2\Sigma_{kj}^{f^\prime,S}(0)\right]\right\},
\label{eq:dv}\\
\delta Z_W&=&-\widetilde{\mbox{Re}}
\left.\frac{\partial\Pi^{WW}(k^2)}{\partial k^2}\right|_{k^2=M_W^2},
\label{eq:dw}\\
\delta Z_{ij}^{f,L}&=&\frac{2}{m_{f,i}^2-m_{f,j}^2}\widetilde{\mbox{Re}}\left[
m_{f,i}^2\Sigma_{ij}^{f,L}\left(m_{f,j}^2\right)
+m_{f,i}m_{f,j}\Sigma_{ij}^{f,R}\left(m_{f,j}^2\right)
\right.\nonumber\\
&&{}+\left.
\left(m_{f,i}^2+m_{f,j}^2\right)\Sigma_{ij}^{f,S}\left(m_{f,j}^2\right)
\right],\qquad i\ne j,
\label{eq:dij}\\
\delta Z_{ii}^{f,L}&=&-\widetilde{\mbox{Re}}\,
\Sigma_{ii}^{f,L}\left(m_{f,i}^2\right)
-m_{f,i}^2\frac{\partial}{\partial p^2}\widetilde{\mbox{Re}}\left[
\Sigma_{ii}^{f,L}(p^2)+\Sigma_{ii}^{f,R}(p^2)+2\Sigma_{ii}^{f,S}(p^2)
\right]_{p^2=m_{f,i}^2}.
\label{eq:dii}
\end{eqnarray}
Here, $\Pi^{BB^\prime}(k^2)$ is the transverse coefficient of the two-point
function, at four-momentum $k$, of the gauge bosons $B$ and $B^\prime$, and
$\Sigma_{ij}^{f,L/R/S}(p^2)$ are the left-handed, right-handed, and scalar
coefficients of the two-point function, at four-momentum $p$, of the fermions
$f_i$ and $f_j$, respectively.
The latter are listed in Eq.~(\ref{eq:fer}) of the Appendix.
The symbol $\widetilde{\mbox{Re}}$ takes the dispersive parts of the loop
integrals appearing in the two-point functions and commutes with
complex-valued parameters, such as $V_{ij}$.
As a consequence, $\delta Z_{ij}^{f,L\dagger}=\delta Z_{ji}^{f,L*}$ is
obtained from $\delta Z_{ji}^{f,L}$ by complex conjugation of the CKM matrix 
elements contained therein.
From Eq.~(\ref{eq:fer}) it hence follows that $\delta Z_{ij}^{f,L\dagger}$
emerges from $\delta Z_{ij}^{f,L}$ through the interchange of $m_{f,i}$ and
$m_{f,j}$.
In particular, we have $\delta Z_{ii}^{f,L\dagger}=\delta Z_{ii}^{f,L}$.
Notice that the vertex correction only depends linearly on $V_{ij}$, which is
factored out in Eq.~(\ref{eq:mol}).\footnote{The last term in Eq.~(9.6) of
Ref.~\cite{DenHabil} should be multiplied with $V_{ij}$.}

All the renormalization constants appearing in Eq.~(\ref{eq:mol}) are
ultraviolet (UV) divergent.
If the renormalized parameters $e$, $s_w$, $V_{ij}$ are to represent physical
observables, they must be gauge independent, and so must the respective
renormalization constants.
This is well established for $\delta e/e$ and $\delta s_w/s_w$ given by
Eqs.~(\ref{eq:de}) and (\ref{eq:ds}), respectively, while an appropriate
definition of $\delta V_{ij}/V_{ij}$, namely Eq.~(\ref{eq:dv}), was proposed
only recently \cite{Mad99,GGM}.
On the other hand, the field renormalization constants in Eq.~(\ref{eq:mol})
are gauge dependent and, with the exception of $\delta Z_{ij}^{f,L}$ with 
$i\ne j$, also infrared (IR) divergent.
Since ${\cal M}_1^+$ and ${\cal M}_2^\pm$ do not yet appear in
Eq.~(\ref{eq:mlo}), the respective form factors are finite and gauge
independent.
However, $\delta F_1^-$ is IR and UV divergent and gauge dependent.
The right-hand side of Eq.~(\ref{eq:mol}) is UV finite and gauge independent
\cite{Mad99,GGM}, but it is IR divergent.
This IR divergence is cancelled in the one-loop expression for the partial
decay width by the real bremsstrahlung correction $\delta_{\rm b}^{\rm ew}$.
Also including the one-loop QCD correction $\delta^{\rm QCD}$, we have
\begin{equation}
\Gamma_1^{Wf_if_j^\prime}=\Gamma_0^{Wf_if_j^\prime}\left(1+\delta^{\rm ew}+
\delta^{\rm QCD}\right),
\label{eq:gol}
\end{equation}
where
\begin{equation}
\delta^{\rm ew}=\delta_{\rm virt}^{\rm ew}+\delta_{\rm b}^{\rm ew},
\label{eq:ew}
\end{equation}
with the virtual electroweak correction
\begin{eqnarray}
\delta_{\rm virt}^{\rm ew}&=&
2\frac{\delta e}{e}-2\frac{\delta s_w}{s_w}
+2\frac{\delta V_{ij}}{V_{ij}}+\delta Z_W
+\sum_k\left(\delta Z_{ik}^{f,L\dagger}V_{kj}
+V_{ik}\delta Z_{kj}^{f^\prime,L}\right)
\nonumber\\
&&{}+
\frac{2}{G_1^-}\sum_{a=1}^{2}\sum_{\sigma=\pm}G_a^\sigma
\delta F_a^\sigma\left(M_W,m_{f,i},m_{f^\prime,j}\right).
\label{eq:vir}
\end{eqnarray}
Note that $\delta_{\rm b}^{\rm ew}$ is gauge independent because
$\Gamma_0^{Wf_if_j^\prime}\delta_{\rm b}^{\rm ew}$ represents the partial
decay width of the physical process $W^+\to f_i\bar f_j^\prime\gamma$ in the
Born approximation.
Thus, $\delta^{\rm ew}$ is finite and gauge independent, as it must be because
it exhausts the leading-order electroweak correction to the physical process
$W^+\to f_i\bar f_j^\prime(\gamma)$.
The QCD correction $\delta^{\rm QCD}$ may be obtained from $\delta^{\rm ew}$
by retaining only the terms containing $Q_f^2$, $Q_{f^\prime}^2$, or
$Q_fQ_{f^\prime}$, setting $Q_f=Q_{f^\prime}=1$, replacing $\alpha$ with the
strong-coupling constant $\alpha_s^{(n_f)}(\mu)$, and including the overall
colour factor $C_F=4/3$.
Also $\delta^{\rm QCD}$ is IR and UV finite and gauge independent, as it must 
be because it exhausts the leading-order QCD correction to the physical
process $W^+\to f_i\bar f_j^\prime(g)$.
In the limit $m_{f,i}=m_{f^\prime,j}=0$, the well-known correction factor
$\delta^{\rm QCD}=1+\alpha_s^{(n_f)}(\mu)/\pi$ is recovered.

In the lepton case, we have $N_C^f=1$ and $\delta^{\rm QCD}=0$.
Furthermore, we have $V_{ij}=\delta_{ij}$ and $\delta V_{ij}=0$ if we assume
the neutrinos to be massless.

We computed all ingredients of Eq.~(\ref{eq:gol}) in $R_\xi$ gauge, with 
arbitrary gauge parameters $\xi_W$, $\xi_Z$, and $\xi_A$, for finite fermion
masses $m_{f,i}$ and $m_{f^\prime,j}$ and general CKM matrix $V_{ij}$.
We regularized the UV divergences by means of dimensional regularization (DR),
in $D=4-2\epsilon$ space-time dimensions, and the IR ones by introducing an 
infinitesimal photon mass $\lambda$.  
To guarantee the correctness of our results, we chose two independent 
approaches.
The first approach was based on the program packages {\tt FeynArts}
\cite{Hah98FA} and {\tt FeynCalc} \cite{MerBoeDen91}, which are written in
{\tt Mathematica}.
{\tt FeynArts} generates the relevant Feynman diagrams and translates them
into $T$-matrix elements, in a format which is readable by {\tt FeynCalc}.
{\tt FeynCalc} then simplifies the expressions and decomposes them into the 
standard one-loop scalar integrals $A_0$, $B_0$, and $C_0$.
The second approach was to essentially perform all the calculations by hand
using well-tested custom-made programs, written in {\tt FORM} \cite{form}, in 
the intermediate steps.
Both approaches led to identical results.
Our results for the gauge-boson self-energies fully agree with those listed in
Eqs.~(7)--(10) of Ref.~\cite{DegSir92} and will not be presented here.
The corresponding formulas in 't~Hooft-Feynman gauge, with
$\xi_W=\xi_Z=\xi_A=1$, may be found in Appendix~B of Ref.~\cite{bak}.
Generic expressions for the fermion self-energies, which were originally
derived in Ref.~\cite{Mad99}, are specified in Eq.~(\ref{eq:fer}) of the
Appendix.
The diagonal fermion wave-function renormalization constants
$\delta Z_{ii}^{f,L}$ suffer from IR divergences.
Therefore, we retained the infinitesimal photon mass $\lambda$ in those parts
of Eq.~(\ref{eq:fer}), from which IR divergences may arise.
In the limit $\lambda=0$, Eq.~(\ref{eq:fer}) agrees with Eqs.~(B1)--(B3) of
Ref.~\cite{GG}.\footnote{The difference may be traced to the terms in
Eq.~(\ref{eq:fer}) that contain
$B_0\left(p^2,\sqrt{\xi_A}\lambda,m_{f,i}\right)$.
If we put $\xi_A=1$ in the argument of this function, then
Eq.~(\ref{eq:fer}) coincides with Eqs.~(B1)--(B3) of Ref.~\cite{GG}.
This difference becomes relevant when IR-divergent quantities such as
$\delta Z_{ii}^{f,L}$ are to be calculated, but it is immaterial for the
purposes of Ref.~\cite{GG}.}
The corresponding formulas in 't~Hooft-Feynman gauge may be found in
Appendix~A of Ref.~\cite{hff}.
We do not display our analytical results for the form factors
$\delta F_a^\sigma$ because they are somewhat lengthy.
They can be compared with the literature in the limiting cases
$\xi_W=\xi_Z=\xi_A=1$ \cite{Den90,DenHabil} or $m_{f,i}=m_{f^\prime,j}=0$
\cite{DegSir92}.
In the first case, we find agreement with Eqs.~(27)--(29) of 
Ref.~\cite{Den90}.\footnote{The function in the seventh line of Eq.~(29)
should carry the superscript ``$\sigma$'' instead of ``$-$''.}
In the second case, only the form factor $\delta F_1^-$ survives, as may be 
seen from Eq.~(\ref{eq:ms}), and we find agreement with Eq.~(33) of
Ref.~\cite{DegSir92}.
We verified the expressions for $\delta_{\rm b}^{\rm ew}$ and
$\delta^{\rm QCD}$ listed in Eqs.~(35) and (37) of Ref.~\cite{Den90},
respectively.

At this point, we should comment on a very recent paper \cite{bar} in which a
new renormalization prescription for the CKM matrix is proposed.
The quantity $T_1$, defined in Eq.~(4) of Ref.~\cite{bar}, corresponds to our
quantity ${\cal M}_1^{Wf_if_j^\prime}$, defined in Eq.~(\ref{eq:mol}) above.
The authors of Ref.~\cite{bar} claim that the finite part of
${\cal M}_1^{Wf_if_j^\prime}$ becomes gauge dependent if $\delta V_{ij}$ is
omitted.
In order to substantiate this claim, they introduce, in Eq.~(23), the
auxiliary quantity $\delta X_{ud}$, which is to represent the difference 
between ${\cal M}_1^{Wf_if_j^\prime}$ and its counterpart for 
$V_{ij}=\delta_{ij}$.
In our notation, this quantity reads
\begin{equation}
\delta X_{ij}=\frac{1}{2}V_{ij}
\left(\delta Z_{ii}^{f,L\dagger}-\delta Z_{ii[1]}^{f,L\dagger}
+\delta Z_{jj}^{f^\prime,L}-\delta Z_{jj[1]}^{f^\prime,L}\right)
+\frac{1}{2}\left(\sum_{k\ne i}\delta Z_{ik}^{f,L\dagger}V_{kj}
+\sum_{k\ne j}V_{ik}\delta Z_{kj}^{f^\prime,L}\right),
\label{eq:xij}
\end{equation}
where the subscript ``$[1]$'' indicates that the identification
$V_{ij}=\delta_{ij}$ is to be made.
They find that this quantity is gauge dependent, and propose to define
$\delta V_{ij}=-\delta X_{ij}$.
From our above discussion, it is clear that $\delta V_{ij}$ must be gauge 
independent, in order for the renormalized parameters $V_{ij}$ to be gauge
independent.
Otherwise, the latter would not qualify as physical observables.
Furthermore, we verified, by inspecting the analytic expressions, that
${\cal M}_1^{Wf_if_j^\prime}
+e{\cal M}_1^-\delta V_{ij}/\left(\sqrt2s_w\right)$
is gauge independent, in accordance with Refs.~\cite{Mad99,GGM}.
This implies that the quantity $\delta X_{ij}$, defined in Eq.~(\ref{eq:xij}),
is also gauge independent, as we explicitly checked.
Finally, we remark that the expression for $T_1$ given in Eq.~(24) of
Ref.~\cite{bar} differs from that given in Eq.~(4) {\it ibidem} by finite
terms because $\delta g/g$ and $\delta Z_W$ do depend on $V_{ij}$.

Equation~(\ref{eq:gol}) is formulated in the pure on-shell renormalization 
scheme, which uses $\alpha$ and the physical particle masses as basic
parameters.
In this scheme, large electroweak corrections arise from fermion loop 
contributions to the renormalizations of $\alpha$ and $s_w$.
As in any charged-current process, these corrections can be greatly reduced by
parameterizing the lowest-order result with Fermi's coupling constant $G_F$
and $M_W$ instead of $\alpha$ and $s_w$.
This can be achieved with the aid of the relationship \cite{sir}
\begin{equation}
G_F=\frac{\pi\alpha}{\sqrt2s_w^2M_W^2}\,\frac{1}{1-\Delta r},
\end{equation}
where $\Delta r$ contains those radiative corrections to the muon decay width
which the SM introduces on top of the purely photonic corrections from within
Fermi's model.
At one loop, we have \cite{sir}
\begin{eqnarray}
\Delta r&=&\frac{\Pi^{WW}(0)-\widetilde{\mbox{Re}}\,
\Pi^{WW}\left(M_W^2\right)}{M_W^2}
+\frac{c_w^2}{s_w^2}\widetilde{\mbox{Re}}\left(
\frac{\Pi^{WW}\left(M_W^2\right)}{M_W^2}
-\frac{\Pi^{ZZ}\left(M_Z^2\right)}{M_Z^2}\right)
+2\frac{c_w}{s_w}\,\frac{\Pi^{ZA}(0)}{M_Z^2}
\nonumber\\
&&{}+\left.\frac{\partial\Pi^{AA}(k^2)}{\partial k^2}\right|_{k^2=0}
+\frac{\alpha}{4\pi s_w^2}\left[\left(\frac{7}{2s_w^2}-2\right)\ln c_w^2+6
\right].
\label{eq:dr}
\end{eqnarray}
The last term herein represents the vertex and box corrections to the muon
decay width in 't~Hooft-Feynman gauge.
Thus, the $\Pi^{BB^\prime}$ functions in Eq.~(\ref{eq:dr}) have to be
evaluated in this gauge, too.
We recall that Eq.~(\ref{eq:dr}) is gauge independent \cite{hem} and finite.
The quantity $\left.\partial\Pi^{AA}(k^2)/\partial k^2\right|_{k^2=0}$ 
receives important contributions from the light-quark flavours, which cannot
be reliably predicted in perturbative QCD.
This problem is usually circumvented by relating the finite and 
gauge-independent quantity
\begin{equation}
\Delta\alpha_{\rm had}^{(5)}=\left[
\left.\frac{\partial\Pi^{AA}(k^2)}{\partial k^2}\right|_{k^2=0}
-\frac{\Pi^{AA}\left(M_Z^2\right)}{M_Z^2}\right]_{udscb}
\end{equation}
via a subtracted dispersion relation to experimental data on the total cross
section of inclusive hadron production in $e^+e^-$ annihilation.
In our numerical analysis, we substitute $\alpha=\sqrt2G_Fs_w^2M_W^2/\pi$ in
Eqs.~(\ref{eq:glo}) and (\ref{eq:ew}) and, in turn, include the term
$-\Delta r$, evaluated from Eq.~(\ref{eq:dr}), within the parentheses on the
right-hand side of Eq.~(\ref{eq:gol}).
Then, the quantity $\left.\partial\Pi^{AA}(k^2)/\partial k^2\right|_{k^2=0}$
exactly cancels, so that the theoretical uncertainty in
$\Delta\alpha_{\rm had}^{(5)}$ does not affect our results.

\boldmath
\section{$\overline{\mathrm{MS}}$ Definition of the CKM Matrix
\label{sec:three}}
\unboldmath

The relationship between the on-shell and $\overline{\mathrm{MS}}$ definitions
of the CKM matrix may be conveniently revealed by considering the identity
\begin{eqnarray}
V_{ij}^0&=&V_{ij}+\delta V_{ij}
\nonumber\\
&=&\bar V_{ij}+\delta\bar V_{ij},
\end{eqnarray}
where the superscript ``0'' labels bare quantities, and 
$\overline{\mathrm{MS}}$ quantities are marked by a bar.
By definition, $\delta\bar V_{ij}$ is the UV-divergent part of
$\delta V_{ij}$, proportional to $1/\epsilon+\ln(4\pi)-\gamma_E$, where
$\gamma_E$ is Euler's constant.
We thus obtain the relationship
\begin{equation}
\bar V_{ij}(\mu)=V_{ij}+\Delta V_{ij}(\mu),
\label{eq:rel}
\end{equation}
where
\begin{equation}
\Delta V_{ij}(\mu)=\delta V_{ij}-\delta \bar{V}_{ij}
\end{equation}
is a finite shift, which depends on the 't~Hooft mass scale $\mu$ of DR.
Inserting Eq.~(\ref{eq:fer}) in Eq.~(\ref{eq:dv}) and performing the
$\overline{\mathrm{MS}}$ subtraction, we find
\begin{eqnarray}
\Delta V_{ij}(\mu)&=&\sum_{k\ne i}\sum_lV_{il}V_{lk}^\dagger V_{kj}
\frac{m_{f,k}^2+m_{f,i}^2}{m_{f,k}^2-m_{f,i}^2}f(m_{f^\prime,l})
\nonumber\\
&&{}+\sum_{k\ne j}\sum_lV_{ik}V_{kl}^\dagger V_{lj}
\frac{m_{f^\prime,k}^2+m_{f^\prime,j}^2}{m_{f^\prime,k}^2-m_{f^\prime,j}^2}
f(m_{f,l}),
\end{eqnarray}
where
\begin{equation}
f(m)=\frac{\alpha}{16\pi s_w^2}\,\frac{x}{2}\left[3\ln\frac{\mu^2}{M_W^2}
-\frac{5x-11}{2(x-1)}+\frac{3x(x-2)}{(x-1)^2}\ln x
\right]_{x=\frac{m^2}{M_W^2}}.
\end{equation}

The $\mu$ dependence of $\bar V_{ij}(\mu)$ is described by the $\beta$
function
\begin{equation}
\beta_{V_{ij}}=\frac{d}{d\ln\mu}\bar V_{ij}(\mu).
\label{eq:bet}
\end{equation}
Inserting Eq.~(\ref{eq:rel}) into Eq.~(\ref{eq:bet}), we obtain the one-loop
expression
\begin{eqnarray}
\beta_{V_{ij}}&=&\frac{3\alpha}{16\pi s_w^2M_W^2}\left[
\sum_{k\ne i}\sum_lV_{il}V_{lk}^\dagger V_{kj}m_{f^\prime,l}^2
\frac{m_{f,k}^2+m_{f,i}^2}{m_{f,k}^2-m_{f,i}^2}
\right.\nonumber\\
&&{}+\left.\sum_{k\ne j}\sum_lV_{ik}V_{kl}^\dagger V_{lj}m_{f,l}^2
\frac{m_{f^\prime,k}^2+m_{f^\prime,j}^2}{m_{f^\prime,k}^2-m_{f^\prime,j}^2}
\right],
\end{eqnarray}
which agrees with the one given in Ref.~\cite{Sas86}.

The standard parameterization of $V_{ij}$ utilizes three angles,
$\theta_{12}$, $\theta_{23}$, and $\theta_{13}$, and one phase, $\delta_{13}$
\cite{Cas98}.
A popular approximation that emphasizes the hierarchy in the size of the 
angles, $\sin\theta_{12}\gg\sin\theta_{23}\gg\sin\theta_{13}$, is due to
Wolfenstein \cite{wol}, where one sets $\lambda=\sin\theta_{12}$, the sine of
the Cabibbo angle, and then writes the other CKM matrix elements in terms of
powers of $\lambda$.
Through $O(\lambda^3)$, one has \cite{Cas98}
\begin{equation}
V=\left(\begin{array}{ccc}
1-\lambda^2/2 & \lambda & A\lambda^3(\rho-i\eta) \\
-\lambda & 1-\lambda^2/2 & A\lambda^2 \\   
A\lambda^3(1-\rho-i\eta) & -A\lambda^2 & 1 
\end{array}\right),
\end{equation}
where $A$, $\rho$, and $\eta$ are real numbers, which were intended to be of 
order unity.
The relationships between the parameters $\lambda$, $A$, $\rho$, and $\eta$ in
the on-shell scheme and their counterparts in the $\overline{\mathrm{MS}}$
scheme may be obtained with the aid of Eq.~(\ref{eq:rel}) and read
\begin{eqnarray}
\bar\lambda&=&\lambda[1+(1-\lambda^2)F(m_u,m_c,m_d,m_s)],
\nonumber\\
\bar A&=&A[1-2F(m_u,m_c,m_d,m_s)+F(m_c,m_t,m_s,m_b)],
\nonumber\\
\bar\rho&=&\rho[1-F(m_u,m_c,m_d,m_s)-F(m_c,m_t,m_s,m_b)+ F(m_u,m_t,m_d,m_b)]
\nonumber\\
&& \mbox{}
+\left(\frac{m_u^2+m_c^2}{m_u^2-m_c^2}-\frac{m_u^2+m_t^2}{m_u^2-m_t^2}\right)
[f(m_d)-f(m_s)]
\nonumber\\
&& \mbox{}
+\left(\frac{m_s^2+m_b^2}{m_s^2-m_b^2}-\frac{m_d^2+m_b^2}{m_d^2-m_b^2}\right)
[f(m_c)-f(m_t)],
\nonumber\\
\bar\eta&=&\eta[1-F(m_u,m_c,m_d,m_s)-F(m_c,m_t,m_s,m_b)+F(m_u,m_t,m_d,m_b)],
\label{eq:wol}
\end{eqnarray}
where
\begin{equation}
F(m_1,m_2,m_3,m_4)=\left\{\frac{m_1^2+m_2^2}{m_1^2-m_2^2}[f(m_3)-f(m_4)]
+\frac{m_3^2+m_4^2}{m_3^2-m_4^2}[f(m_1)-f(m_2)]\right\}.
\end{equation}
Exploiting the strong hierarchy among the up- and down-type quark masses,
which satisfy $m_u\ll m_c\ll m_t$ and $m_d\ll m_s\ll m_b$, respectively, and
the significant mass splittings within the second and third quark generations,
$m_s\ll m_c$ and $m_b\ll m_t$, respectively, we can derive the following
approximation formulas:
\begin{eqnarray}
\bar\lambda&=&\lambda\left[1+(1-\lambda^2)f(m_c)\right],
\nonumber\\
\bar A&=&A\left[1+f(m_t)\right],
\nonumber\\
\bar\rho&=&\rho,
\nonumber\\
\bar\eta&=&\eta.
\end{eqnarray}
For $\mu=M_W$, these approximations agree with the exact results, evaluated
from Eq.~(\ref{eq:wol}), within an error of less than $10^{-5}$.

\section{Numerical Results\label{sec:four}}

We now describe our numerical analysis of Eq.~(\ref{eq:gol}), with the 
modifications specified at the end of Sec.~\ref{sec:two}.
We evaluated the $A_0$, $B_0$, and $C_0$ functions with the aid of the
{\tt Fortran} program package {\tt FF} \cite{OldVer90}, which is embedded into
the {\tt Mathematica} environment through the program package {\tt LoopTools}
\cite{Hah98LT}.
We performed several checks for the correctness and the stability of our
numerical results.
As we demonstrated in Sec.~\ref{sec:two}, Eq.~(\ref{eq:gol}) is IR and UV
finite and gauge independent, as a consequence of cancellations among the
various terms contained in Eqs.~(\ref{eq:ew}) and (\ref{eq:vir}).
We can check the IR finiteness and gauge independence numerically by varying
the IR regulator $\lambda$ and the gauge parameters $\xi_W$, $\xi_Z$, and
$\xi_A$, respectively.
In the physical limit $D\to4$, the UV divergences appear as terms proportional
to $1/\epsilon+\ln(4\pi)-\gamma_E$, which are accompanied by a term
$\ln(\mu^2/M^2)$, with $M$ being a characteristic mass scale of the considered
loop integral.
Although we nullified $1/\epsilon+\ln(4\pi)-\gamma_E$ in our computer program,
we can check the UV finiteness numerically by varying the 't~Hooft mass scale
$\mu$.
A further check on the stability of our numerical analysis can be obtained by
directly evaluating the two- and three-point tensor integrals with the aid of
{\tt LoopTools} instead of applying the Passarino-Veltman reduction 
algorithm~\cite{pas}.
Our numerical analysis passed all these checks.
Finally, we managed to reproduce the numerical results of Ref.~\cite{DenHabil}
after adopting the definition of $\delta V_{ij}$, the choice of gauge, and the
values of the input parameters from there.

We use the following input parameters \cite{Cas98,ewwg}:
\begin{eqnarray}
\begin{array}{lll}
\alpha=1/137.03599976, &
G_F=1.16639\times10^{-5}~\mbox{GeV}^{-2},\quad &
\alpha_s^{(5)}(M_Z)=0.1181, \\
M_W=80.419~\mbox{GeV}, &
M_Z=91.1871~\mbox{GeV}, &
\\
m_e=0.510998902~\mbox{MeV},\quad &
m_\mu=105.658357~\mbox{MeV}, &
m_\tau=1777.03~\mbox{MeV}, \\
m_u=3~\mbox{MeV}, &
m_d=6~\mbox{MeV}, &
m_s=123~\mbox{MeV}, \\
m_c=1.5~\mbox{GeV}, &
m_b=4.6~\mbox{GeV}, &
m_t=174.3~\mbox{GeV}, \\
s_{12}=0.223, &
s_{23}=0.040, &
s_{13}=0.004.
\label{eq:inp}
\end{array}
\end{eqnarray}
Here, $m_u$, $m_d$, and $m_s$ correspond to current-quark masses, and $m_c$,
$m_b$, and $m_t$ to pole masses.
Since the contributions from the light quarks $q=u,d,s,c,b$ come with the
suppression factors $m_q^2/M_W^2$, the considerable uncertainties in the 
values of $m_q$ do not jeopardize the reliability of our theoretical
predictions.
We take the neutrinos to be massless.
For simplicity, we assume that $\delta_{13}=0$.
Then, the values for $s_{ij}=\sin\theta_{ij}$ provided in the last row of
Eq.~(\ref{eq:inp}) lead to
\begin{eqnarray}
\begin{array}{lll}
V_{ud}=0.975, & V_{us}=0.223, & V_{ub}=0.004, \\
V_{cd}=-0.223,\quad & V_{cs}=0.974, & V_{cb}=0.040, \\
V_{td}=0.005, & V_{ts}=-0.040,\quad & V_{tb}=0.999.  
\label{eq:ckm}
\end{array}
\end{eqnarray}
These values approximately satisfy the unitarity condition
$V_{ik}V_{kj}^\dagger=\delta_{ij}$.\footnote{Notice that the indices of
$V_{ij}$ refer to generations rather than quark flavours.
For example, we have $V_{12}=V_{us}$, $V_{21}=V_{cd}$, and
$V_{12}^\dagger=V_{21}^*=V_{cd}^*$.}
We evaluate $\alpha_s^{(n_f)}(\mu)$ appearing in $\delta^{\rm QCD}$ at the 
renormalization scale $\mu=M_W$ with $n_f=5$ active quark flavours from the
one-loop relation
\begin{equation}
\alpha_s^{(n_f)}(M_W)=\frac{\alpha_s^{(n_f)}(M_Z)}
{1+\alpha_s^{(n_f)}(M_Z)\beta_0\ln(M_W^2/M_Z^2)/\pi},
\end{equation}
where $\beta_0=11/4-n_f/6$.
For the Higgs-boson mass, we consider the values $M_H=100$, 250, and 600~GeV.

\begin{table}
\begin{center}
\begin{tabular}{|l|l|l|}
\hline\hline
partial width & Ref.~\cite{Den90a} & Refs.~\cite{Mad99,GGM} \\
\hline
$\Gamma(W\to ud)$ &0.67122024  & 0.67122024 \\
$\Gamma(W\to us)\times 10$ &0.35125894  & 0.35125894 \\
$\Gamma(W\to ub)\times 10^4$ &0.11273928  & 0.11273992   \\
$\Gamma(W\to cd)\times 10$ &0.35113436  & 0.35113436    \\
$\Gamma(W\to cs)$ &0.67001209  & 0.67001209     \\
$\Gamma(W\to cb)\times 10^2$ &0.11279438  &   0.11279502    \\
$\Gamma(W\to\mbox{hadrons})$ &1.41261088 & 1.41261088 \\
\hline\hline
\end{tabular}
\caption{\label{tab:one} Partial widths (in GeV) of the hadronic $W$-boson
decays at one loop, for $M_H=250$~GeV.
The results obtained in 't~Hooft-Feynman gauge with $\delta V_{ij}$ as defined
in Ref.~\protect\cite{Den90a} are compared with ours.}
\end{center}
\end{table}

We now present our numerical results.
We first investigate the quantitative significance of the definition of
$\delta V_{ij}$.
Toward this end, we compare, in Table~\ref{tab:one}, our results for the
partial widths of the various hadronic $W$-boson decay channels with those
obtained in 't~Hooft-Feynman gauge with the definition of $\delta V_{ij}$
proposed in Ref.~\cite{Den90a}, assuming $M_H=250$~GeV.
The relative deviations are largest for the final states involving the $b$
quark, where they are of order $\alpha m_b^2/(\pi M_W^2)\approx10^{-5}$.
Although small against the present experimental accuracies \cite{Cas98,ewwg},
they are of the same order as the entire shifts due to the renormalization of
the CKM matrix \cite{Den90a}.
We stress that the numbers in the second column of Table~\ref{tab:one} do 
depend on the choice of gauge.
However, this gauge dependence turns out to be feeble.
In Table~\ref{tab:two}, we present our tree-level and one-loop results for the
leptonic and hadronic partial decay widths of the $W$ boson, assuming 
$M_H=100$, 250, or 600~GeV.
In the leptonic channels, the radiative corrections are nearly flavour 
independent and amount to approximately $-0.3\%$.
In the hadronic channels, the corrections range between 3.5\% and 3.8\% and 
are dominantly of QCD origin.
In all cases, the $M_H$ dependence is feeble, of relative order $10^{-5}$.
Finally, we determine the uncertainties in our theoretical prediction for the
total $W$-boson decay width $\Gamma_W$ due to the errors on our input
parameters.
Specifically, the variations of
$G_F$, $\alpha_s^{(5)}(M_Z)$, $M_W$, $M_Z$,
$m_c$, $m_b$, $m_t$,
$s_{12}$, $s_{23}$, and $s_{13}$ by
$\pm1\times10^{-10}~\mbox{GeV}^{-2}$, $\pm0.002$, $\pm38$~MeV, $\pm2.1$~MeV,
$\pm0.1$~GeV, $\pm0.2$~GeV, $\pm5.1$~GeV,
$\pm0.004$, $\pm0.003$, and $\pm0.002$ \cite{Cas98,ewwg} shift $\Gamma_W$ by
$\pm0.02$, $\pm0.9$, $\pm3.0$, $\pm7.5\times10^{-4}$,
$\pm0.02$, $\pm1\times10^{-4}$, $\pm0.033$,
$\pm3\times10^{-5}$, $\pm3\times10^{-4}$, and $\pm3\times10^{-5}$~MeV,
respectively.
The residual parametric uncertainties are marginal.

\begin{table}
\begin{center}
\begin{tabular}{|l|l|l|l|l|}
\hline\hline
partial width & Born & \multicolumn{3}{c|}{one loop} \\
\cline{3-5}
& & $M_H=100$~GeV & $M_H=250$~GeV & $M_H=600$~GeV \\
\hline
$\Gamma(W\to e\nu_e)$ &0.2275641  &0.2268336 &0.2268444 &0.2268458 \\
$\Gamma(W\to\mu\nu_\mu)$ &0.2275635  &0.2268331  &0.2268439  & 0.2268452   \\
$\Gamma(W\to\tau\nu_\tau)$ &0.2273974 &0.2266717  &0.2266823  &0.2266836  \\
$\Gamma(W\to\mbox{leptons})$ &0.6825249 &0.6803384  &0.6803706 &0.6803746 \\
\hline
$\Gamma(W\to ud)$ &0.6487322 & 0.6711894  &0.6712202  &0.6712242    \\
$\Gamma(W\to us)\times 10$ &0.3394894 &0.3512428   &0.3512589  &0.3512610  \\
$\Gamma(W\to ub)\times 10^4$ &0.1086947 & 0.1127355 &0.1127399 &0.1127402  \\
$\Gamma(W\to cd)\times 10$ &0.3392503 &0.3511185   &0.3511344 & 0.3511363 \\
$\Gamma(W\to cs)$ &0.6473169 & 0.6699818  & 0.6700121  &0.6700158  \\
$\Gamma(W\to cb)\times 10^2$ &0.1086359 &0.1127907   & 0.1127950 &0.1127953 \\
$\Gamma(W\to\mbox{hadrons})$ &1.3650203 &1.4125466 &1.4126109  &1.4126190 \\
\hline
$\Gamma_W$ &2.0475452  &2.0928850 &2.0929814  &2.0929936   \\
\hline\hline
\end{tabular}
\caption{\label{tab:two} Partial decay widths (in GeV) of the $W$ boson at
tree level and at one loop, for $M_H=100$, 250, or 600~GeV.}
\end{center}
\end{table}

\section{Conclusions\label{sec:five}}

We calculated the partial decay widths of the $W$ boson at one loop in the SM
using the on-shell scheme endowed with the gauge-independent definition of the
CKM matrix $V_{ij}$ recently proposed in Refs.~\cite{Mad99,GGM}.
Working in $R_\xi$ gauge, we explicitly verified that the final expressions
are independent of the gauge parameters.
In particular, the renormalization constant $\delta V_{ij}/V_{ij}$ 
(\ref{eq:dv}) and the one-loop amplitude ${\cal M}_1^{Wf_if_j^\prime}$
(\ref{eq:mol}) of the $W$-boson decay to quarks~(\ref{eq:pro}) with this
renormalization constant removed are separately gauge independent.
In this respect, we disagree with the findings of Ref.~\cite{bar}.
The difference between our analysis and the corresponding one with the
gauge-dependent definition of $\delta V_{ij}/V_{ij}$ from Ref.~\cite{Den90a} 
is of the same order as the entire effect due to the renormalization of the
CKM matrix, but it is small compared to the present experimental precision.
Furthermore, we established the relationship between the on-shell and
$\overline{\mathrm{MS}}$ definitions of the CKM matrix, both in its generic
form \cite{Cas98} and in the Wolfenstein parameterization \cite{wol}.
As a by-product of our analysis, we recovered the beta function of the CKM
matrix \cite{Sas86}.

\vspace{1cm}
\begin{center}
{\bf Note added}
\end{center}
\smallskip

In the meantime, a revised version of Ref.~\cite{bar} has appeared, in which
the weaknesses pointed out in Sec.~\ref{sec:two} have been remedied.

\vspace{1cm}
\begin{center}
{\bf Acknowledgements}
\end{center}
\smallskip

We are grateful to K.-P. O. Diener for checking our results for the triangle 
diagrams and for verifying the expressions for $\delta_{\rm b}^{\rm ew}$ and
$\delta^{\rm QCD}$ listed in Eqs.~(35) and (37) of Ref.~\cite{Den90},
respectively.
We enjoyed fruitful discussions with P. A. Grassi. 
M. S. would like to thank T. Hahn and G. Weiglein for valuable advice
concerning {\tt FeynArts} and {\tt FeynCalc}.
This work was supported in part by the Deutsche Forschungsgemeinschaft through
Grant No.\ KN~365/1-1, by the Bundesministerium f\"ur Bildung und Forschung
through Grant No.\ 05~HT9GUA~3, and by the European Commission through the
Research Training Network {\it Quantum Chromodynamics and the Deep Structure
of Elementary Particles} under Contract No.\ ERBFMRX-CT98-0194.

\renewcommand {\theequation}{\Alph{section}.\arabic{equation}}
\begin{appendix}

\setcounter{equation}{0} 
\section{Fermion Two-Point Functions}

The unrenormalized two-point function describing the fermion transition
$f_j\to f_i$, at four-momentum $p$, may be decomposed as
\begin{equation}
\Gamma_{ij}^f(p)=i\left[\delta_{ij}\left(\not\!p-m_{f,i}\right)
+\not\!p\omega_-\Sigma_{ij}^{f,L}(p^2)+\not\!p\omega_+\Sigma_{ij}^{f,R}(p^2)
+\left(m_{f,i}\omega_-+m_{f,j}\omega_+\right)\Sigma_{ij}^{f,S}(p^2)\right].
\end{equation}
At one loop in $R_\xi$ gauge, we find for the coefficient functions herein
\begin{eqnarray}
\Sigma_{ij}^{f,L}(p^2)&=&-\frac{\alpha}{4\pi}\left\{
\delta_{ij}Q_f^2\left[2B_1\left(p^2,m_{f,i},\lambda\right)+1 
  +B_0\left(p^2,\lambda,m_{f,i}\right)
  -\xi_A B_0\left(p^2,\sqrt{\xi_A}\lambda,m_{f,i}\right)
\right.\right.\nonumber\\
&&{}+\left.\frac{p^2-m_{f,i}^2}{\lambda^2}\left(
   B_1\left(p^2,\lambda,m_{f,i}\right)
  -B_1\left(p^2,\sqrt{\xi_A}\lambda,m_{f,i}\right)
  \right)
  \right]
\nonumber\\
&&{}+\delta_{ij}\left(g_f^-\right)^2\left[
2B_1\left(p^2,m_{f,i},M_Z\right)+1\right.
\nonumber\\
&&{}+B_0\left(p^2,m_{f,i},M_Z\right)
-\xi_ZB_0\left(p^2,m_{f,i},\sqrt{\xi_Z}M_Z\right)
\nonumber\\
&&{}+\left.\frac{p^2-m_{f,i}^2}{M_Z^2}\left(B_1\left(p^2,M_Z,m_{f,i}\right)
-B_1\left(p^2,\sqrt{\xi_Z}M_Z,m_{f,i}\right)\right)\right]
\nonumber\\
&&{}+\delta_{ij}\frac{m_{f,i}^2}{4s_w^2M_W^2}\left[
B_1\left(p^2,m_{f,i},\sqrt{\xi_Z}M_Z\right)+B_1\left(p^2,m_{f,i},M_H\right)
\right]
\nonumber\\
&&{}+\frac{1}{2s_w^2}\sum_{k}V_{ik}V_{kj}^\dagger\left[
2B_1\left(p^2,m_{f^\prime,k},M_W\right)
+\frac{m_{f^\prime,k}^2}{M_W^2}
B_1\left(p^2,m_{f^\prime,k},\sqrt{\xi_W}M_W\right)\right.
\nonumber\\
&&{}+1+B_0\left(p^2,m_{f^\prime,k},M_W\right)
-\xi_WB_0\left(p^2,m_{f^\prime,k},\sqrt{\xi_W}M_W\right)
\nonumber\\
&&{}+\left.\frac{p^2-m_{f^\prime,k}^2}{M_W^2}
\left[B_1\left(p^2,M_W,m_{f^\prime,k}\right)
-B_1\left(p^2,\sqrt{\xi_W}M_W,m_{f^\prime,k}\right)\right]\right\},
\nonumber\\
\Sigma_{ij}^{f,R}(p^2)&=&-\frac{\alpha}{4\pi}\left\{
\delta_{ij}Q_f^2\left[2B_1\left(p^2,m_{f,i},\lambda\right)+1 
  +B_0\left(p^2,\lambda,m_{f,i}\right)
  -\xi_A B_0\left(p^2,\sqrt{\xi_A}\lambda,m_{f,i}\right)
\right.\right.\nonumber\\
&&{}+\left.\frac{p^2-m_{f,i}^2}{\lambda^2}\left(
   B_1\left(p^2,\lambda,m_{f,i}\right)
  -B_1\left(p^2,\sqrt{\xi_A}\lambda,m_{f,i}\right)
  \right)
  \right]
\nonumber\\
&&{}+\delta_{ij}\left(g_f^+\right)^2\left[
2B_1\left(p^2,m_{f,i},M_Z\right)+1\right.
\nonumber\\
&&{}+B_0\left(p^2,m_{f,i},M_Z\right)
-\xi_ZB_0\left(p^2,m_{f,i},\sqrt{\xi_Z}M_Z\right)
\nonumber\\
&&{}+\left.\frac{p^2-m_{f,i}^2}{M_Z^2}\left(B_1\left(p^2,M_Z,m_{f,i}\right)
-B_1\left(p^2,\sqrt{\xi_Z}M_Z,m_{f,i}\right)\right)\right]
\nonumber\\
&&{}+\delta_{ij}\frac{m_{f,i}^2}{4s_w^2M_W^2}\left[
B_1\left(p^2,m_{f,i},\sqrt{\xi_Z}M_Z\right)+B_1\left(p^2,m_{f,i},M_H\right)
\right]
\nonumber\\
&&{}+\left.\frac{m_{f,i}m_{f^\prime,j}}{2s_w^2M_W^2}
\sum_{k}V_{ik}V_{kj}^\dagger
B_1\left(p^2,m_{f^\prime,k},\sqrt{\xi_W}M_W\right)\right\},
\nonumber\\
\Sigma_{ij}^{f,S}(p^2)&=&-\frac{\alpha}{4\pi}\left\{
\delta_{ij}Q_f^2\left[4B_0\left(p^2,m_{f,i},\lambda\right)-2 
-B_0\left(p^2,m_{f,i},\lambda\right)
+\xi_A B_0\left(p^2,m_{f,i},\sqrt{\xi_A}\lambda\right)\right]
\right.\nonumber\\
&&{}+\delta_{ij}g_f^+g_f^-\left[
4B_0\left(p^2,m_{f,i},M_Z\right)-2\right.
\nonumber\\
&&{}-\left.B_0\left(p^2,m_{f,i},M_Z\right)
+\xi_ZB_0\left(p^2,m_{f,i},\sqrt{\xi_Z}M_Z\right)\right]
\nonumber\\
&&{}+\delta_{ij}\frac{m_{f,i}^2}{4s_w^2M_W^2}\left[
B_0\left(p^2,m_{f,i},\sqrt{\xi_Z}M_Z\right)-B_0\left(p^2,m_{f,i},M_H\right)
\right]
\nonumber\\
&&{}+\left.\frac{1}{2s_w^2M_W^2}
\sum_{k}V_{ik}V_{kj}^\dagger m_{f^\prime,k}^2
B_0\left(p^2,m_{f^\prime,k},\sqrt{\xi_W}M_W\right)\right\},
\label{eq:fer}
\end{eqnarray}
where
\begin{eqnarray}
B_0\left(p^2,m_0,m_1\right)&=&\frac{(2\pi\mu)^{4-D}}{i\pi^2}
\int\frac{d^Dq}{\left(q^2-m_0^2\right)\left[(q+p)^2-m_1^2\right]},
\\
B_1\left(p^2,m_0,m_1\right)&=&\frac{m_1^2-m_0^2}{2p^2}
\left[B_0\left(p^2,m_0,m_1\right)-B_0(0,m_0,m_1)\right]
-\frac{1}{2}B_0\left(p^2,m_0,m_1\right).\nonumber
\end{eqnarray}

\end{appendix}

\end{document}